\documentclass[a4paper]{article}

\pdfoutput=1

\usepackage[T1]{fontenc} 
\usepackage{lmodern} 
\usepackage[defblank]{paralist} 
\usepackage[obeyspaces]{url} 

\usepackage{specs}
\usepackage{mymacros}
\usepackage{tikz}
\usetikzlibrary{shapes,arrows}
\usetikzlibrary{positioning,fit}
\tikzstyle{blockstandard} = [rectangle, draw=black, thick, fill=gray!30, 
text centered, rounded corners]
\tikzstyle{syncstandard} = [double arrow, draw=red, thick, fill=red!20, 
draw,font=\tiny, double arrow head extend=0.05cm,inner ysep=.005cm]
\tikzstyle{block} = [blockstandard,text width=1.7cm, minimum height=0.7cm,node distance=1.2cm]
\tikzstyle{blockleft} = [block,node distance=2cm]
\tikzstyle{blockbelow} = [block,node distance=1.4cm]
\tikzstyle{sync}=[syncstandard,inner xsep=1.2cm]

\tikzstyle{block3} = [blockstandard,text width=3.7cm, node distance=1.2cm]
\tikzstyle{block3left} = [block3,node distance=4cm]
\tikzstyle{block3below} = [block3,node distance=1.4cm]
\tikzstyle{sync3}=[syncstandard,inner xsep=3.3cm]

\newcommand{\EXTRAS}[1]{}

\newcommand{\OLDMODEL}[1]{}

\usepackage{RR}
    \RRprojet{Regal}
    \RRtheme{\THCom}
    \URRocq

\begin{document}

\RRNo{8083}
\RRauthor{
  Annette Bieniusa, {\footnotesize INRIA \& UPMC, Paris, France}
  \\
  Marek Zawirski, {\footnotesize INRIA \& UPMC, Paris, France}
  \\
  Nuno Pregui\c{c}a, {\footnotesize CITI, Universidade Nova de Lisboa,
    Portugal}
  \\
  Marc Shapiro, {\footnotesize INRIA \& LIP6, Paris, France}
  \\
  Carlos Baquero, {\footnotesize HASLab, INESC TEC \& Universidade do Minho, Portugal}
  \\
  Valter Balegas, {\footnotesize CITI, Universidade Nova de Lisboa, Portugal}
  \\
  S\'{e}rgio  Duarte {\footnotesize CITI, Universidade Nova de Lisboa, Portugal}
}
\authorhead{Bieniusa, Zawirski, Preguiça, Shapiro, Baquero, Balegas, Duarte}

\RRdate{Octobre 2012}
\RRtitle{Optimisation d'un type de données ensemble répliqué sans conflit}
\titlehead{An optimized conflict-free replicated set}
\RRetitle{
	An Optimized Conflict-free Replicated Set
  \thanks{
This research is supported in part by ANR project
   \href{http://concordant.lip6.fr/}{ConcoRDanT} (ANR-10-BLAN 0208),
by ERDF, COMPETE Programme, by Google European Doctoral Fellowship
in Distributed Computing received by Marek Zawirski, and FCT projects
\#PTDC/EIA-EIA/104022/2008 and \#PTDC/EIA-EIA/108963/2008. }
}

\RRabstract{
  Eventual consistency of replicated data supports concurrent
  updates, reduces latency and improves fault tolerance, but forgoes
  strong consistency.
  Accordingly, several cloud computing platforms implement
  eventually-consistent data types.

  The set is a widespread and useful abstraction, and many replicated set
  designs have been proposed.
  We present a reasoning abstraction, \emph{permutation equivalence}, that
  systematizes the characterization of the expected concurrency semantics
  of concurrent types.
  Under this framework we present one of the existing conflict-free
  replicated data types, Observed-Remove Set.

  Furthermore, in order to decrease the size of meta-data, we propose
  a new optimization to avoid tombstones.
  This approach that can be transposed  to other data types, such as
  maps, graphs or sequences.
}

\RRresume{
  La réplication des données avec cohérence à terme permet les mises à
  jour concurrentes, réduit la latence, et améliore la tolérance aux
  fautes, mais abandonne la cohérence forte.
  Aussi, cette approche est utilisée dans plusieurs plateformes de
  nuage.

  L'\emph{ensemble} (Set) est une abstraction largement utilisée, et
  plusieurs modèles d'ensemble répliqués ont été proposés.
  Nous présentons \emph{l'équivalence de permutation}, un principe de
  raisonnement qui caractérise de façon systématique la sémantique
  attendue d'un type de données concurrent.
  Ce principe nous permet d'expliquer la conception un type déjà connu,
  \emph{Observed-Remove Set}.

  Par ailleurs, afin de diminuer la taille des méta-données, nous
  proposons une nouvelle optimisation qui évite les « pierres
  tombales ».
  Cette approche peut se transposer à d'autres types de données, comme
  les mappes, les graphes ou les séquences.
}

\RRmotcle{Réplication des données, réplication optimiste, opérations commutatives} 
\RRkeyword{Data replication, optimistic replication, commutative operations}

\makeRR

\section{Introduction}

Eventual consistency of
replicated data supports concurrent updates, reduces latency and improves
fault tolerance, but forgoes strong consistency (e.g., linearisability).
Accordingly, several cloud computing platforms implement
eventually-consistent replicated sets
\cite{app:rep:optim:1606,rep:syn:1661}.
%
%
%
Eventual Consistency, allows concurrent
updates at different replicas, under the expectation that replicas will
eventually converge \cite{rep:syn:pan:1624}.
However, solutions for addressing concurrent updates tend to be either
limited or very complex and error-prone \cite{optim:rep:syn:1500}.

We follow a different approach: 
Strong Eventual Consistency (SEC) \cite{syn:rep:sh143} 
requires a deterministic outcome for any pair of
concurrent updates. Thus, different replicas can be updated in parallel,
and concurrent updates are resolved locally, without requiring consensus.
Some simple conditions (e.g., that concurrent updates commute with
one another) are sufficient to ensure SEC\@.
Data types that satisfy these conditions are called Conflict-Free
Replicated Data Types (CRDTs).
Replicas of a CRDT object can be updated without synchronization and
are guaranteed to converge.
This approach has been adopted in several works
\cite{app:rep:optim:1501,db:rep:1651,alg:rep:sh131,app:rep:1652,rep:syn:1661}.

The set is a pervasive data type, used either directly or as a component of
more complex data types, such as maps or graphs.  
This paper highlights the semantics of sets under eventual consistency, and
introduces an optimized set implementation, \emph{Optimized Observed Remove
Set}.

\section{Principle of Permutation Equivalence}
\label{sec:permutation-eq}

The sequential semantics of a set are well known, and are defined by
individual updates, e.g., $\{ \true \} \add(e) \{ e \in S \}$ (in
``\{pre-condition\} computation \{post-condition\}'' notation), where
$S$ denotes its abstract state.
However, the semantics of concurrent modifications is left underspecified or implementation-driven.




We propose the following \emph{Principle of Permutation Equivalence}
\cite{BZPSBBD12} 
to express that concurrent behaviour conforms to the
sequential specification:
\emph{``If all sequential permutations of updates lead to equivalent states,
then it should also hold that concurrent executions of the updates lead to
equivalent states.'' }
It implies the following behavior, for some updates $u$ and $u'$: $$\{P\}
u;u' \{Q\} \land \{P\} u';u \{Q'\} \land Q \Leftrightarrow Q' \;\; \implies
\;\; \{P\} u \parallel u' \{Q\}$$

Specifically for replicated sets, the Principle of Permutation Equivalence requires that 
 $\{ e \ne f \} \add(e) \parallel \remove(f)  \{ e \in S \land f \notin S \}$, and similarly for operations on different elements or idempotent operations.
Only the pair $\add(e) \parallel \remove(e)$ is unspecified by the
principle, since $\add(e); \remove(e)$ differs from $\remove(e);
\add(e)$.
Any of the following post-conditions ensures a deterministic result:
{\small \[
\begin{array}{@{}cl@{}}
  \{ \bottom_{e} \in S \} &  \text{-- Error mark}\\
  \{ e \in S \} & \text{-- \add wins} \\
  \{ e \notin S \} & \text{-- \remove wins} \\
  \{ \add(e) >_{\CLK} \remove(e) \iff e \in S \} & \text{-- Last
  Writer Wins (LWW)}\\
\end{array}
\] }\\[-1ex]
where $<_{\CLK}$ compares unique clocks associated with the
operations.
Note that not all concurrency semantics can be explained as a sequential
permutation; for instance no sequential execution ever results in an
error mark.



\begin{figure*}[tb]
\begin{minipage}{\textwidth}
\subfigure[Dynamo shopping cart\label{fig:amazon:anomaly}]{
\includegraphics[width=0.26\textwidth]{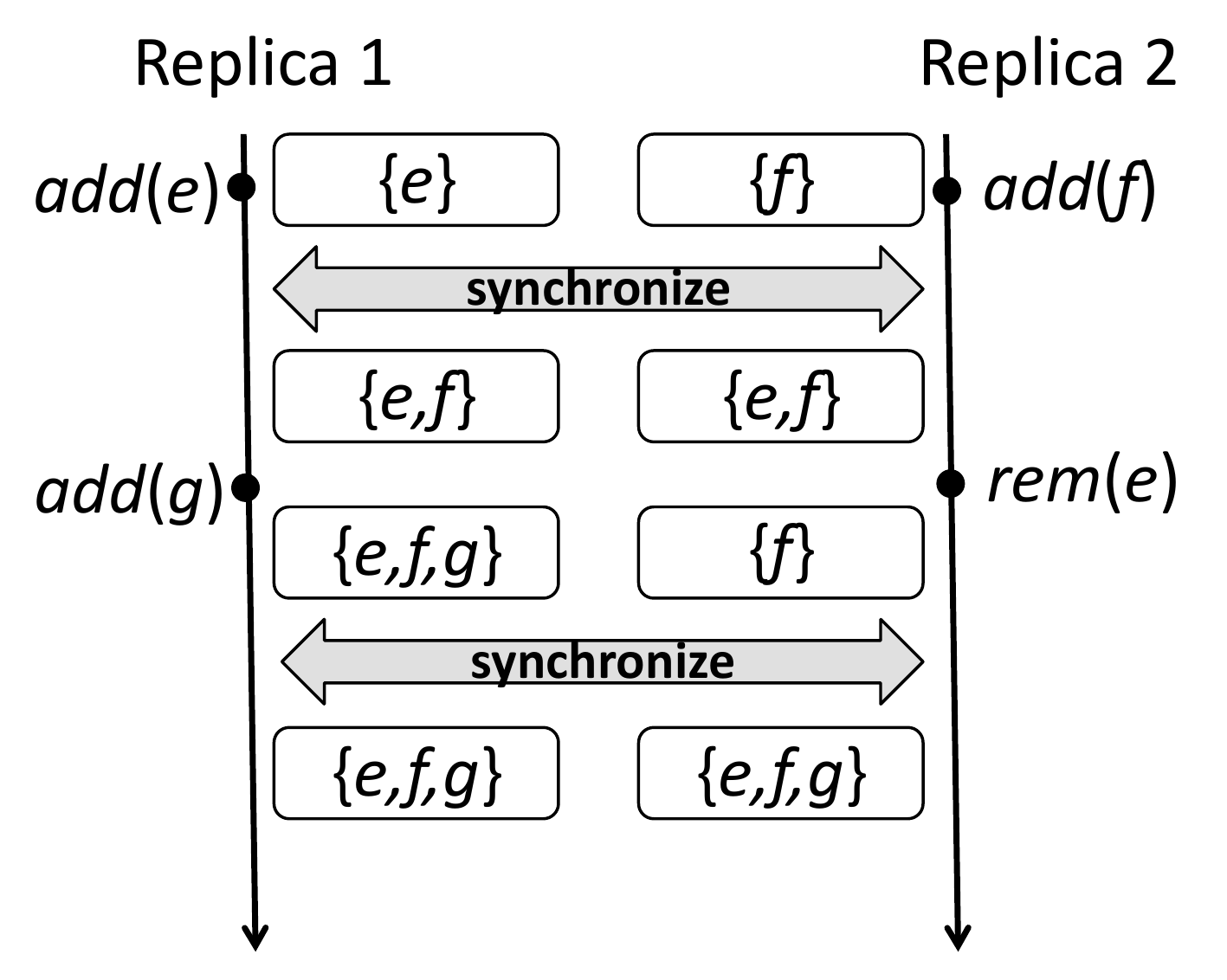}
}
\subfigure[C-Set\label{fig:cset:anomaly}]{
\includegraphics[width=0.26\textwidth]{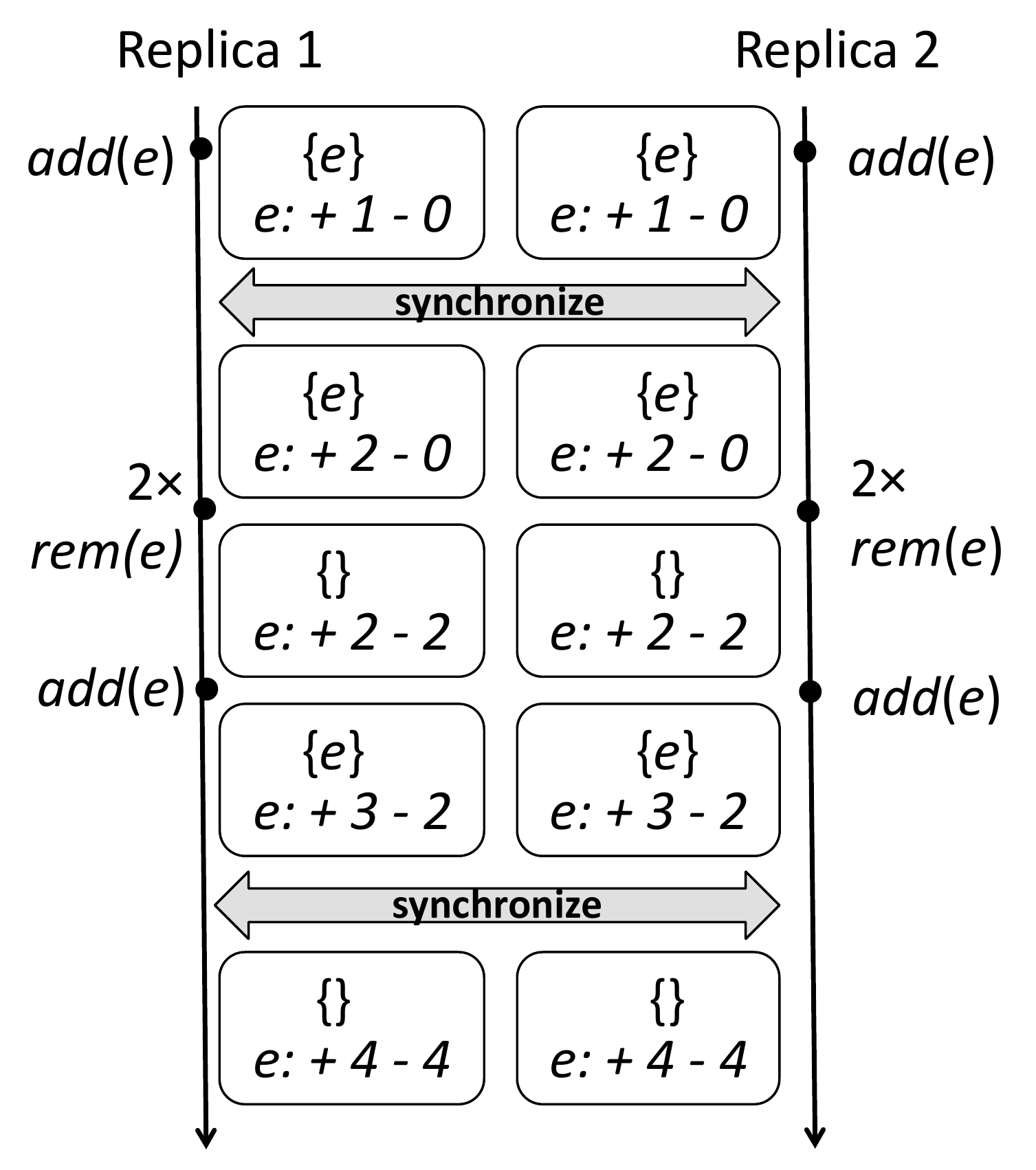}
}
\subfigure[OR-Set\label{fig:cset:orset:anomaly}]{
\includegraphics[width=0.45\textwidth]{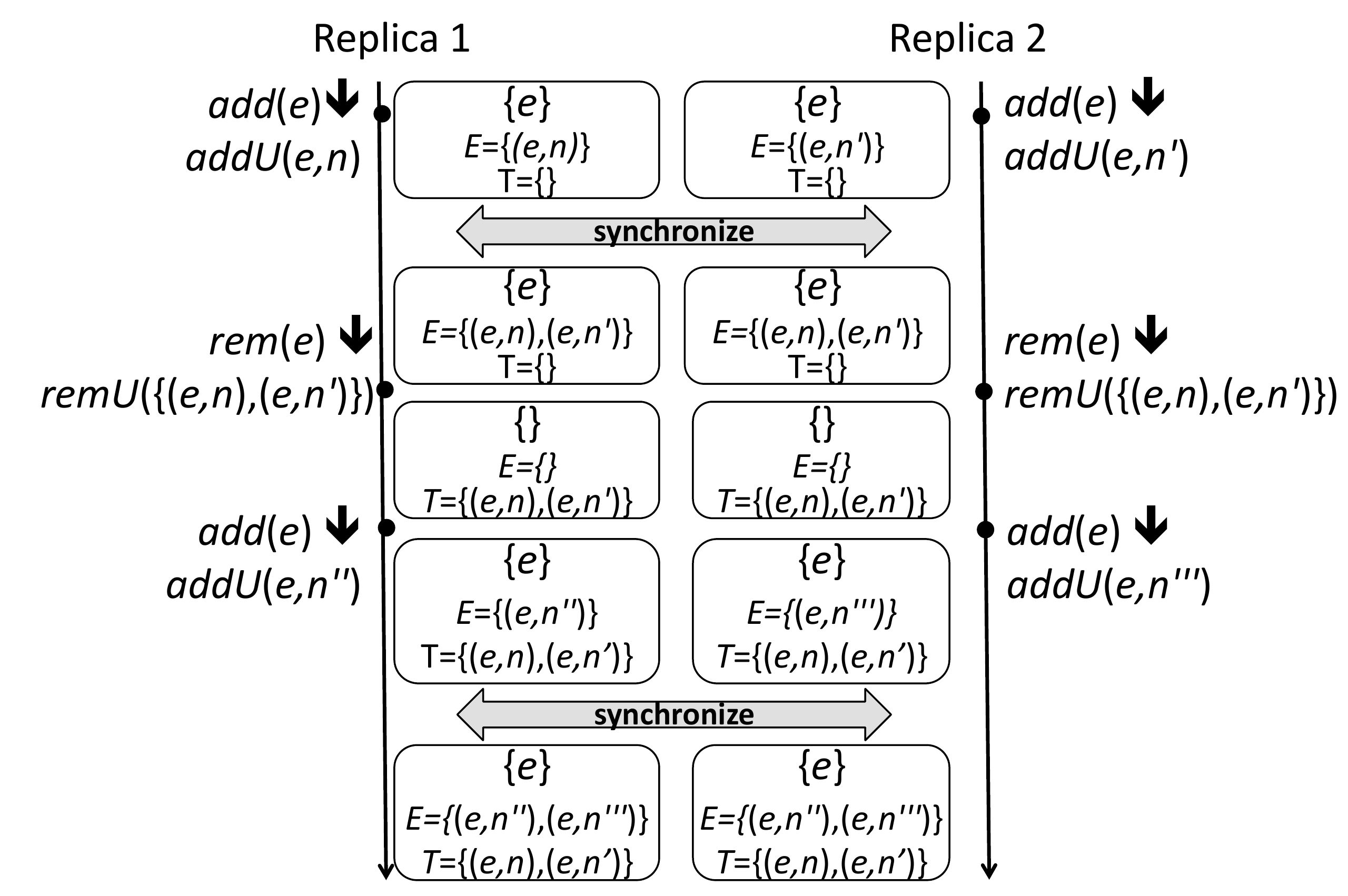}
}
\caption{Examples of anomalies and a correct design.}
\label{fig:anomaly}
\end{minipage}
\end{figure*}

\section{A Review of Existing Replicated Set Designs}
\label{sec:anomalies}
In the past, several designs have been proposed for maintaining 
a replicated set.
Most of them violate the Principle of Permutation Equivalence
(Fig.~\ref{fig:anomaly}).
For instance, the Amazon Dynamo shopping cart
\cite{app:rep:optim:1606} is implemented using a register
supporting \Read and \Write (assignment) operations, offering the
standard sequential semantics.
When two \Write{}s occur concurrently, the next \Read returns their
union.
As noted by the authors themselves, in case of concurrent updates even
on unrelated elements, a \remove may be undone
(Fig.~\ref{fig:amazon:anomaly}).

Sovran et al.\ and Asian et al.\
\cite{rep:syn:1661,c-set2011} propose a set variant,
{C-Set},
where for each element the associated \add and \remove updates are counted.
The element is in the abstraction if their difference is positive.
C-Set violates the Principle of Permutation Equivalence
(Fig.~\ref{fig:cset:anomaly}).
When delivering the updates to both replicas as sketched, the add and remove counts are equal, i.e., $e$ is not in the abstraction, even though the last update at each replica is $\add(e)$.


\section{Add-wins Replicated Sets}

In Section \ref{sec:permutation-eq} we have shown that when considering
concurrent \add and \remove operations over the same element, one among
several post-conditions can be chosen. Considering the case of an \add wins
semantics we now recall \cite{syn:rep:sh143} 
the CRDT design of an
Observed Remove Set, or \emph{OR-Set}, and then introduce an optimized
design that preserves the \emph{OR-Set} behaviour and greatly improves its
space complexity. 

These CRDT specifications follow a new notation with mixed state- and
operation-based update propagation. Although the formalization of this mixed
model, and the associated proof obligations that check compliance to CRDT
requisites, is out of the scope of this report the notation is easy to infer
from the standard CRDT model 
\cite{syn:rep:sh143,rep:syn:sh138,syn:sh144}.


{\bf System model synopsis:} We consider a single object, replicated at a
given set of processes/replicas. A client of the object may invoke an
operation at some replica of its choice, which is called the \emph{source}
of the operation.  A query executes entirely at the source.  An update
applies its side effects first to the source replica, then (eventually) at
all replicas, in the \emph{downstream} for that update. To this effect, an
update is modeled as an \emph{update pair} $(p,u)$ that includes two
operations such that $p$ is a side-effect free \emph{prepare(-update)}
operation and $u$ is an \emph{effect(-update)} operation; the source
executes the prepare and effect atomically; downstream replicas execute only
the effect $u$. In the mixed state- and operation-based modelling, replica
state can both be changed by applying an effect operation or by merging
state from another replica of the same object. The monotonic evolution of
replica states is described by a compare operation, supplied with each CRDT
specification.

\subsection{Observed Remove Set}

\specone{OR-Set: Add-wins replicated set\label{spec:addwins:set}}{
\begin{algorithmic}
  \Payload{set $E$, set $T$}    \Comment{$E$: elements; $T$: tombstones}
           \State{}  \Comment{sets of pairs \emph{\{ ({element} $e$, {unique-tag} $n$), \ldots \}}}
    \Initial{$\emptyset, \emptyset$}
    \Query{\contains}{element $e$}{boolean b}
        \Let{$b = (\exists n: (e, n) \in E)$}
        \EndQuery
    \Query{\elements}{}{set $S$}
      \Let{$S = \{e | \exists n: (e,n) \in E\}$}
    \EndQuery
    \Update{\add}{element $e$}
        \AtSource{$e$}
            \Let{$n{} = \unique()$}\Comment{$\unique()$ returns a unique tag}
            \EndAtSource
        \Downstream{$e, n$}
            \State{$E := E \union \{ (e, n) \} \setminus T$} \Comment{$e$ + unique tag}
            \EndDownstream
        \EndUpdate
    \Update{\remove}{element $e$}
        \AtSource{$e$} \Comment{Collect pairs containing $e$}
            \Let{$R = \{ (e, n) | \exists n: (e,n) \in E \} $}
            \EndAtSource
        \Downstream{$R$}            \Comment{Remove pairs observed at source}
            \State{$E := E \setminus R$}
            \State{$T := T \union R$}
            \EndDownstream
        \EndUpdate
    \LEL{$A$}{$B$}
        \Let{$b = ((A.E \union A.T) \subseteq (B.E \union B.T)) \land (A.T \subseteq B.T)$}
    \MergeML{$B$}
        \State{$E := (E \setminus B.T) \union (B.E \setminus T)$}
        \State{$T := T \union B.T$}
        \EndMergeML
  \end{algorithmic}
  }

Figure~\ref{spec:addwins:set} shows our specification for an add-wins
replicated set CRDT\@.
Its concurrent specification 
$ \{ P \}  u_{0} \parallel \ldots{} \parallel u_{n-1}  \{ Q \}$
is for each element $e$ defined as follows:
\begin{compactitem}
\item
  $\forall i, u_{i} = \remove(e) \implies Q = (e \notin S)$
\item
  $\exists i: u_{i} = \add(e) \implies Q = (e \in S)$.
\end{compactitem}

To implement add-wins, the idea is to distinguish different invocations of
$\add(e)$ by adding a hidden unique token $n$, and effectively store $(e,
n)$ pair.  A pair $(e, n)$ is removed by adding it to a tombstone set.  An
element can be always added again, because the new pair $(e, n')$ uses
always a fresh token, different from the old one, $n' \ne n$.  If the same
element $e$ is both added and removed concurrently, the update-prepare of
\remove concerns only observed pairs $(e, n_1), (e, n_2), \ldots$ and not
the concurrently-added unique pair $(e, n')$.  Therefore the \add wins by
adding a new pair.  We call this object an Observed Remove Set, or
\emph{OR-Set}.  As illustrated in Figure~\ref{fig:cset:orset:anomaly},
\emph{OR-Set} is immune from the anomaly that plagues {C-Set}.

{\bf Space complexity:} The payload size of \emph{OR-Set} is at any moment
bounded by the number of all applied add (\emph{effect-update}) operations.

\subsection{Optimized Observed Remove Set}

\begin{figure}[h!]
\caption{Optimized OR-Set (Opt-OR-Set).\label{spec:addwins:set:notombs}}
\begin{algorithmic}
  \Payload{set $E$, vect $v$}  \Comment{$E$: elements, set of triples \emph{
           ({element} $e$, {timestamp} $c$, replica $i$)}}
           \State    \Comment{$v$: summary (vector) of received triples}
    \Initial{$\emptyset, [0, \ldots, 0]$}
    \Query{\contains}{element $e$}{boolean b}
        \Let{$b = (\exists c, i: (e, c, i) \in E)$}
        \EndQuery
    \Query{\elements}{}{set $S$}
      \Let{$S = \{e | \exists c, i: (e,c,i) \in E\}$}
    \EndQuery
    \Update{\add}{element $e$}
        \AtSource{$e$}
            \Let{$r{} = \myID()$}\Comment{$r$ = source replica}
            \Let{$c{} = v[r] + 1$}
            \EndAtSource
        \Downstream{$e, c, r$}
            \Pre{causal delivery}
            \If{$c > v[r]$}
                \Let{$O = \{ (e, c', r) \in E| c' < c\}$}
                \State{$v[r] := c$}
                \State{$E := E \union \{ (e, c, r) \} \setminus O$}
                \EndIf
            \EndDownstream
        \EndUpdate
    \Update{\remove}{element $e$}
        \AtSource{$e$}\Comment{Collect all unique triples containing $e$}
            \Let{$R = \{ (e, c, i) \in E \} $}
            \EndAtSource
        \Downstream{$R$} \Comment{Remove triples observed at source}
            \Pre{causal delivery}
            \State{$E := E \setminus R$} 
            \EndDownstream
        \EndUpdate
    \LEML{$A$}{$B$}
      \Let{$R = \{ (c, i) | 0 < c \leq A.v[i]\; \land \not\exists e: (e,c,i) \in A.E \}$}
    	\Let{$R' = \{ (c, i) | 0 < c \leq B.v[i]\; \land \not\exists e: (e,c,i) \in B.E \}$} 
        \Let{$b = A.v \leq B.v \land R \subseteq R'$}
        \EndLEML
    \MergeML{$B$}
        \Let{$M{} = (E \inter B.E)$}
        \Let{$M' = \{(e,c,i) \in E \setminus B.E | c > B.v[i] \}$}
        \Let{$M'' = \{(e,c,i) \in B.E \setminus E | c > v[i] \}$}
        \Let{$U = M{} \union M' \union M''$}
        \Let{$O = \{(e, c, i) \in U| \exists (e, c', i) \in U: c < c'\}$}
        \State{$E := U \setminus O$}
        \State{$v := [\max(v[0], B.v[0]), \ldots, \max(v[n], B.v[n])]$}
        \EndMergeML
  \end{algorithmic}
  \end{figure}

The \emph{OR-Set} design uses extensively unique identifiers and tombstones,
as other CRDTs \cite{alg:rep:sh131,app:rep:1652,rep:syn:sh138}.  We now show
how to make CRDT practical by minimizing the required meta-data.

{\bf Immediately discarding tombstones:} When comparing two payloads $P$ and
$P'$, respectively containing some element $e$ and the other not, it is
important to know if $e$ has been recently added to $P$, or if it was
recently removed from $P'$.  The presented add-wins set uses tombstones to
unambiguously answer this question, even when updates are delivered out of
order or multiple times.

Tombstones accumulate (as a consequence of the monotonic
semilattice requirement); if they cannot be discarded, memory
requirements grow with the number of operations.
To address this issue, Wuu's 2P-Set \cite{app:rep:optim:1501}
garbage-collects tombstones that have been delivered everywhere, basically by
waiting for acknowledgements from each process to every other process.
This adds communication and processing overhead, and requires all
processes to be correct. 
We devise a novel technique to eliminate tombstones without these
limitations and offer conflict-free semantics at an affordable cost.
We present our solution using add-wins as the example.

To recapitulate, in \emph{OR-Set}, adding an element $e$ creates a new unique
$(e,n)$ pair to the $E$ part of the payload.  Removing an element moves all
pairs containing $e$ observed at the source from $E$ to $T$.%
\footnote{
  A  practical implementation will just set a mark bit on the representation
of the removed pair and will deallocate any other associated storage.
Consider for instance the extension of \emph{OR-Set} to a map: a key will have some
associated value, e.g., $E$ would contain triples $(e,n,\mathit{value})$.
When the key is removed, $\mathit{value}$ can be discarded, but the
corresponding $(e,n)$ pair(s) must remain in $T$.  }
Note that adding some pair $(e,n)$ always happens-before removing the same
pair $(e,n)$.  If updates are delivered only in causal order,
once, the \add always executes before any related \remove{}s, and the
tombstone set $T$ is not necessary when executing operations.  However, we
also need to support state-based \merge, which joins two replicas possibly
unrelated by happens-before.  When merging two replicas in which only one
replica has some pair $(e,n)$, we need to know if the pair has been added to
the replica that contains it or if it was removed in the other replica.

We leverage these observations to propose a novel \remove algorithm that
discards a removed pair immediately and works safely with \merge.  It
compactly records happens-before information to summarizes removed
elements.  Figure~\ref{spec:addwins:set:notombs} presents \emph{Optimized OR-Set}
(Opt-OR-Set) based on this approach.

Each replica $i$ maintains a vector $v$ \cite{Parker83Detection} to
summarize the unique identifiers it has already observed.  Entry $v[j] = n$
at replica $i$ indicates that this replica has observed $n$ successive
identifiers generated at $j$: $(1,j),(2,j),\ldots,(n,j)$.  Replica $i$
maintains its local counter as the $i$-th entry in the vector $v[i]$,
initially $0$.  A replica generates new unique identifiers $(c,i)$ by
incrementing its local counter.  Note that to summarize successive
identifiers in a  vector, OptORSet requires causal delivery of
updates.\footnote{ It is easy to extend this solution for updates delivered
out of happens-before order by using instead a version vector with
exceptions \cite{DBLP:journals/dc/MalkhiT07}.}

When \add is invoked, the source associates it with a unique identifier made
of the next local counter value and source replica identifier.  When the
\add is delivered to a downstream replica, it should have an effect only if
it has not been previously delivered; for this, it checks if the unique
identifier is incorporated in the downstream replica's vector.  When
\merge{}ing payloads, an element should be in the merged state only if:
either it is in both payloads (set $M$ in
Figure~\ref{spec:addwins:set:notombs}), or it is in the local payload and
not recently removed from the remote one (set $M'$) or vice-versa ($M''$) -
an element has been removed if it is not in the payload but its identifier
is reflected in the replica's vector.

This approach can be generalized to any CRDT where elements are added
and removed, e.g., a sequence \cite{alg:rep:sh131,app:rep:1652} or a
graph \cite{syn:sh144}.


{\bf Coalescing repeated adds:} Another source of memory growth in the
original \emph{OR-Set} is due to the elements added several times.  Similarly to
tombstones, they pollute the state with unique identifiers for every \add{}.
We observe that for every combination of element and source replica, it is
enough to keep the identifier of the latest \add, which subsumes previously
added elements.  The OptORSet specification leverages this observation in
\add{} and \merge{} definitions, by discarding unnecessary identifiers (set
$O$).

{\bf Space complexity:} The payload size of \mbox{OptORSet} set is bounded
by $O(|\elements| n + n)$ at any moment, where $n$ is the number of
processes in the systems and $|\elements|$ is the number of elements present
in the set.  The first component corresponds to the maximum number of
timestamps in set $E$ and the second captures the size of the vector $v$.
In the common case, where the number of processes repeatedly invoking
\add{}s can be considered a constant, the payload size is $O(|\elements| +
n)$.

\section{Conclusions}

Conflict-Free Replicated Data Types (CRDTs) allow a system to maintain
multiple replicas of data that are updated without requiring synchronization
while guaranteeing Strong Eventual Consistency.  This allows, for example, a
cloud infrastructure to maintain replicas of data in data centers spread over
large geographical distance and still provide low access latency by choosing
the closest, to client, data center.

In this paper we reviewed existing replicated set designs and contrasted then
with the CRDT \emph{OR-Set} design, under the principle of permutation
equivalence. Having in mind that the base \emph{OR-Set} favored simplicity at the
expense of scalability, we introduced a new optimized design, \emph{Optimized
OR-Set}, that greatly improves its scalability and should favor efficient
implementations of sets and other CRDTs that share the \emph{OR-Set} design
techniques.

{
\footnotesize
\newcommand{\textcommabelow}[1]{\c{#1}}
\bibliographystyle{plain}
\bibliography{bib,shapiro-bib,local}
}

\end{document}